\documentclass{article}
\usepackage{spconf,amsmath,graphicx,hyperref}
\usepackage{booktabs} 
\usepackage{amssymb}
\usepackage{colortbl}
\usepackage{graphicx}
\usepackage{url}
\usepackage{cite}
\definecolor{conditionrow}{RGB}{235,242,250}
\definecolor{checkblue}{RGB}{31,93,153}
\newcommand{\cmark}{\ensuremath{\checkmark}}

\newcommand{\cdcs}{\textsc{CDCS}}

\newcommand{\cdcsFive}{\textsc{CDCS-5}}
\newcommand{\csg}{\textsc{CSG}}
\newcommand{\ud}{\textsc{UD}}
\newcommand{\sims}{\operatorname{cos}}
\newcommand{\fspk}{F_{\mathrm{spk}}}

\title{Speaker-Consistent Grounding for Confusion-Resilient LLM-Based Target Speech Extraction}
\name{%
\shortstack{
Hanyu Meng$^{1}$, Hao Shi$^{2}$, Zhanhong He$^{3}$,
Qiquan Zhang$^{1}$, Yang Xiao$^{4}$,\\
Sungwoo Cho$^{5}$, Haowei Lou$^{1}$, Haizhou Li$^{6}$
}
}

\address{%
$^{1}$ UNSW Sydney, Australia 
$^{2}$ Kyoto University, Japan 
$^{3}$ University of Western Australia, Australia \\
$^{4}$ University of Melbourne, Australia 
$^{5}$ KAIST AI, Republic of Korea
$^{6}$ CUHK-Shenzhen, China
}
\begin{document}
\ninept
\maketitle
\begin{abstract}
Deterministic target-speaker extraction (TSE) tends to preserve observed linguistic content but may extract the interfering speaker. In contrast, autoregressive generative TSE improves perceptual quality, but is prone to content hallucination and speaker identity drift, which can be mitigated by model grounding with the evidence from the deterministic TSE. However, grounding such a generative TSE model with speaker-confused evidence may therefore reinforce the error. To address this gap, we proposed a repair-before-grounding framework with a training-free evidence-repair stage. Conditioning-diverse candidate selection (CDCS) selects target speaker-consistent evidence from complementary frozen TSE outputs. The selected waveform can then be used by code-space grounding (CSG) to constrain LLM-based TSE decoding. On Libri2Mix, CDCS reduces wrong-speaker extraction errors from 6.87\% to 0.75\% on clean mixtures and from 18.95\% to 10.46\% under noise. With the LLM-based TSE checkpoint unchanged, CSG further reduces noisy content- and acoustic-switch rates from 12.79\% and 1.68\% to 9.10\% and 0.51\%, respectively. Despite this stronger grounding, generative TSE retains its predicted perceptual-quality advantage over direct deterministic extraction. These results show that repairing evidence identity before grounding improves the reliability of generative TSE.\footnote{Audio demos and code are available at \url{https://hanyu-meng.github.io/LLM-TSE-Grounding-Demo/}.}
\end{abstract}

\vspace{-1.5mm}
\section{Introduction}
\label{sec:intro}
\vspace{-1.5mm}
Target-speaker extraction (TSE) aims to recover the speech of a target speaker from an overlapping mixture using an auxiliary enrollment utterance, potentially in the presence of background noise~\cite{zmolikova2023overview}. Representative systems operate in either the spectrogram or waveform domain and condition extraction on an enrollment-derived speaker representation~\cite{zmolikova2019speakerbeam,ge2020spexplus}. However, enrollment conditioning does not guarantee correct speaker selection. Beyond residual distortion or leakage, an extractor may instead follow the competing speaker, a failure termed ``target confusion''~\cite{zhao2022confusion,liu2023xsepformer}. Such an output can remain clear and intelligible while carrying the interferer's identity and linguistic content. Perceptual quality alone therefore cannot certify reliable target extraction.

Diffusion and speech-token TSE systems exploit learned speech priors to reconstruct high-quality speech and can improve perceptual quality over direct extraction~\cite{kamo2023diffusion,tang2024tselm,tang2025lauratse,yu2024generation}. However, preserving linguistic content remains a challenge: autoregressive language models may generate fluent content unsupported by the acoustic input~\cite{zeng2026discgen}. Evidence-guided decoding can mitigate such hallucinations by constraining generation toward a deterministic speech estimate, as in code-space grounding (CSG) for autoregressive speech enhancement~\cite{shi2026grounded}. In TSE, consistency with an extracted estimate does not guarantee correct speaker selection: grounding may preserve the interferer's speech if target confusion has already occurred. Conversely, selecting target-consistent evidence does not itself constrain subsequent generative drift. We therefore aim to retain the quality benefits of generative TSE while addressing two complementary requirements: evidence identity before generation to mitigate target confusion, and decoding consistency during generation to limit hallucination.

To improve target evidence, existing work explores enrollment selection, contextual conditioning, and explicit target/interferer
rectification~\cite{sato2022enrolment,yang2024context,wang2024tcr}. Other inference-time methods iteratively generate and score estimates or update enrollment using reliable session history~\cite{you2026multistep,liu2026evotse}. We instead investigate whether a fixed-budget, non-iterative selection over complementary frozen extraction paths can repair speaker-confused evidence before generative decoding.

\begin{figure*}[t!]
    \centering
    \includegraphics[width=0.98\textwidth]{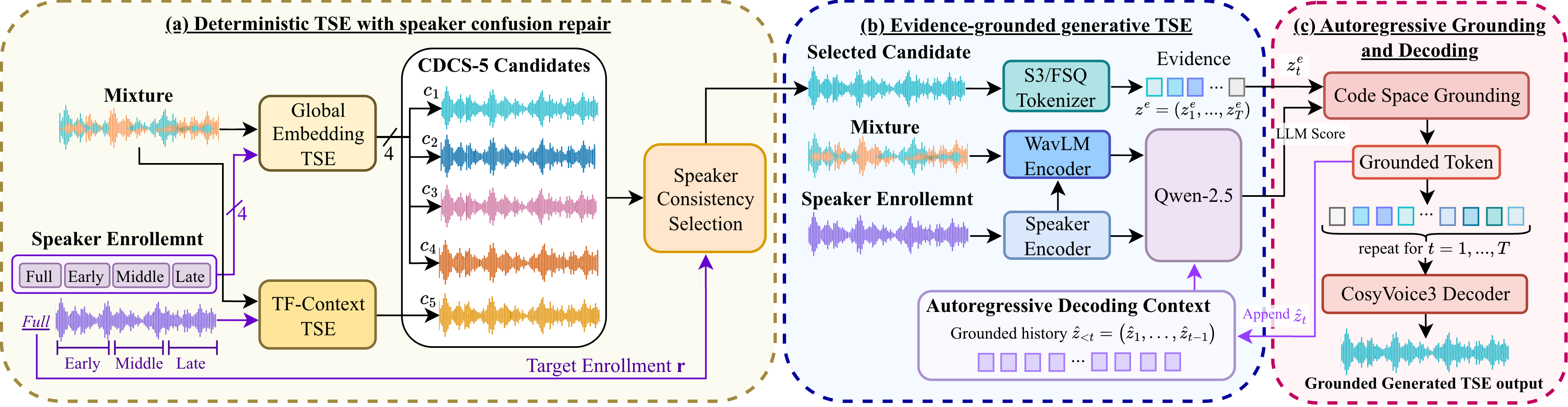}
    \vspace{-2mm}
    \caption{Overview of the proposed repair-before-grounding framework.
    (a) CDCS-5 selects target-speaker-consistent evidence from five frozen
    deterministic TSE estimates. The selected waveform is either returned
    directly or (b) tokenized as evidence together with mixture and
    target-speaker conditioning. (c) CSG uses this evidence to constrain
    autoregressive Qwen-TSE decoding.}
    \vspace{-5mm}
    \label{fig:method}
\end{figure*}

We address this question with conditioning-diverse candidate selection (\cdcs{}), an inference-time evidence-repair mechanism, and integrate it with fixed and adaptive CSG in the repair-before-grounding framework shown in Fig.~\ref{fig:method}. The main contributions are as follows:
\begingroup
\setlength{\leftmargini}{1.2em}
\setlength{\labelsep}{0.4em}
\begin{itemize}
\setlength{\itemsep}{0.15em}
\setlength{\parsep}{0pt}
\setlength{\parskip}{0pt}
\item \textbf{Training-free evidence repair.} We propose \cdcsFive{}, which constructs five deterministic candidates using complementary enrollment views and two frozen TSE paths. A frozen speaker encoder selects the candidate most consistent with the target enrollment.
\item \textbf{Speaker- and content-consistent grounding.} Our repair-before-grounding framework combines discriminative evidence with generative reconstruction: \cdcs{} mitigates target confusion, while fixed and adaptive CSG constrain a frozen Qwen-TSE checkpoint to limit content drift.
\item \textbf{Reduced confusion with high-quality reconstruction.} On Libri2Mix, \cdcsFive{} reduces wrong-speaker extraction with few new errors on initially correct controls. Under noise, grounding with the selected evidence reduces content switching relative to ungrounded decoding while retaining higher DNSMOS and UTMOS scores than those of the deterministic TSE baselines.
\end{itemize}
\endgroup
\vspace{-4mm}
\section{Speaker-Consistent Grounding}\label{sec:method}
\vspace{-1.5mm}
\subsection{System Overview}
Figure~\ref{fig:method} summarizes the proposed two-stage framework. Since grounding an incorrect estimate may preserve the interferer's speech, CDCS first constructs five complementary estimates using frozen deterministic TSE paths and selects the candidate most consistent with the target enrollment. The selected waveform, $c^\ast$, is either returned directly or tokenized into an evidence sequence. In the generative TSE branch, code-space grounding (\csg{}) uses this sequence to constrain the fixed Qwen-TSE decoder. This repair-before-grounding design first mitigates speaker confusion in the extracted evidence and then limits autoregressive content drift during generation.
\vspace{-2mm}
\subsection{Conditioning-diverse candidate selection (CDCS)}\label{sec:cdcs}
\vspace{-1mm}
Let $y=s_{\mathrm{tar}}+s_{\mathrm{int}}+n$ denote a two-speaker mixture containing target speech $s_{\mathrm{tar}}$, interfering speech $s_{\mathrm{int}}$, and noise $n$, where $n=0$ for clean mixtures. Given a target enrollment $r$, a speech extractor returns an estimate $E(y,r)$, which may incorrectly extract the interferer. To mitigate target confusion, \cdcs{} constructs the candidate set $\mathcal{C}(y,r)$:
\begin{equation}
\mathcal{C}(y,r)=
\{E_{\mathrm{spk}}(y,\rho_k(r))\}_{k=0}^{3}
\cup
\{E_{\mathrm{tf}}(y,r)\},
\label{eq:candidates}
\end{equation}
where $E_{\mathrm{spk}}$ denotes the frozen non-causal TSE-BSRNN
extractor implemented in WeSep~\cite{wang2024wesep}. It is conditioned on a global embedding of the target speaker enrollment. We define $\rho_0(r)=r$ and divide the enrollment into three temporal thirds. From each third, we select the maximum-energy window of duration $T_{\mathrm{view}}(r)=\min(\tau,T_r/3)$, where $\tau$ is a hyperparameter controlling the maximum duration of each local enrollment view, and $T_r$ denotes the duration of enrollment $r$. Windows are evaluated with a 10-ms hop, with ties resolved by the earliest start time.

The fifth candidate, $E_{\mathrm{tf}}(y,r)$, uses the same non-causal BSRNN backbone
but replaces the pooled global speaker embedding with a mixture-aligned enrollment time-frequency (TF) map and cross-attended frame-level ECAPA features. This preserves fine-grained enrollment cues and helps recover cases in which the global-embedding candidates confuse the target speaker.

Next, \cdcs{} uses a frozen ECAPA-TDNN speaker encoder $\fspk$~\cite{desplanques2020ecapa} to score each candidate $c_i\in\mathcal{C}(y,r)$ by its cosine similarity to the full target enrollment and selects the highest-scoring candidate:
\vspace{-2mm}
\begin{equation}
q_i=\sims\!\left(\fspk(c_i),\fspk(r)\right),\;
i^*=\operatorname*{arg\,max}_{i} q_i,\;
c^*=c_{i^*}.
\label{eq:selection}
\vspace{-3mm}
\end{equation}
If multiple candidates obtain the same score, we select the first in the fixed order full, early, middle, late, and the TF-context enrollment. As $\mathcal{C}(y,r)$ contains five candidates, we refer to this configuration as \cdcsFive{} throughout the remainder of the paper.
\vspace{-2mm}
\subsection{Evidence-Grounded LLM-based TSE}\label{sec:grounding}
\vspace{-1mm}
The selected waveform $c^\ast$ is converted into an evidence token sequence $z^e$ using the 25-Hz supervised semantic speech (S3) tokenizer from CosyVoice3~\cite{du2025cosyvoice3}, which employs finite scalar quantization (FSQ)~\cite{mentzer2024fsq}. For autoregressive speech-token generation, we initialize the LLM backbone from Qwen2.5-0.5B-Instruct~\cite{qwen2024} and fine-tune all its parameters. The enrollment-derived speaker embedding conditions the WavLM encoder~\cite{chen2022wavlm} during mixture encoding and is also supplied directly to Qwen-TSE together with the resulting speaker-conditioned features. The Qwen-TSE checkpoint remains fixed during inference, while CSG uses $z^e$ to constrain token selection.
\begin{table*}[t!]
\centering
\caption{Deterministic and generative TSE under clean and noisy conditions. WER and switch rates are percentages; DNSMOS denotes P.808. BSRNN-TSE is the full-enrolment baseline. Best and second-best values per condition are bold and underlined, respectively.}
\vspace{1mm}
\label{tab:overall}
\definecolor{conditionrowsoft}{RGB}{240,244,248}

\fontsize{8pt}{9pt}\selectfont
\setlength{\tabcolsep}{4.2pt}
\renewcommand{\arraystretch}{1.04}

\begin{tabular}{@{}lcccccccc@{}}
\toprule
\textbf{System}
& \textbf{Grounding}
& \shortstack{\textbf{Confusion Repair}}
& \textbf{WER}$\downarrow$
& \shortstack{\textbf{Content Switch}$\downarrow$}
& \shortstack{\textbf{Acoustic Switch}$\downarrow$}
& \shortstack{\textbf{Spk. margin}$\uparrow$}
& \textbf{DNSMOS}$\uparrow$
& \textbf{UTMOS}$\uparrow$
\\
\midrule

\rowcolor{conditionrowsoft}[0pt][0pt]
\multicolumn{9}{@{}l@{}}{%
\rule{0pt}{2.2ex}\hspace{2pt}\textit{Clean TEST}%
}
\\[-0.3mm]

BSRNN-TSE~\cite{yu2023bsrnn}
& --
& --
& 11.41
& 7.00
& 6.98
& \underline{0.449}
& 3.688
& 3.606
\\

BSRNN-TSE-CDCS-5
& --
& \cmark
& \textbf{7.02}
& \textbf{0.88}
& 0.78
& \textbf{0.506}
& 3.699
& 3.647
\\

Qwen-TSE-UD
& --
& --
& 8.99
& 0.92
& 0.43
& 0.421
& \underline{3.705}
& \underline{3.913}
\\

Qwen-TSE-fixed-CSG
& \cmark
& \cmark
& \underline{8.04}
& 0.92
& \underline{0.23}
& 0.422
& 3.699
& 3.910
\\

Qwen-TSE-adaptive-CSG
& \cmark
& \cmark
& 8.12
& \underline{0.90}
& \textbf{0.20}
& 0.421
& \textbf{3.711}
& \textbf{3.921}
\\

\addlinespace[0.15mm]
\midrule

\rowcolor{conditionrowsoft}[0pt][0pt]
\multicolumn{9}{@{}l@{}}{%
\rule{0pt}{2.2ex}\hspace{2pt}\textit{Noisy TEST}%
}
\\[-0.3mm]

BSRNN-TSE~\cite{yu2023bsrnn}
& --
& --
& \underline{39.85}
& 18.79
& 18.23
& 0.266
& 3.031
& 2.001
\\

BSRNN-TSE-CDCS-5
& --
& \cmark
& \textbf{30.67}
& 11.02
& 10.39
& 0.327
& 2.980
& 1.948
\\

Qwen-TSE-UD
& --
& --
& 46.70
& 12.79
& 1.68
& 0.377
& 3.494
& \underline{3.216}
\\

Qwen-TSE-fixed-CSG
& \cmark
& \cmark
& 42.11
& \textbf{9.10}
& \textbf{0.51}
& \underline{0.386}
& \underline{3.498}
& 3.147
\\

Qwen-TSE-adaptive-CSG
& \cmark
& \cmark
& 42.73
& \underline{9.37}
& \underline{0.53}
& \textbf{0.388}
& \textbf{3.503}
& \textbf{3.253}
\\

\bottomrule
\end{tabular}

\vspace{-5mm}
\end{table*}

Under ungrounded decoding (\ud{}), Qwen-TSE selects the highest-probability token at each decoding step $t$. In contrast, \csg{}~\cite{shi2026grounded} penalizes token hypotheses that disagree with the repaired evidence $z^e$, thereby reducing content drift and hallucination. To tolerate minor temporal misalignment, we define the distance from a candidate next token $v\in\mathcal{V}$ to the local evidence around step $t$ as
\begin{equation}
D_{H,w}(v,z_t^e)
=
\min_{\delta\in\{-w,\ldots,w\}}
D_H\!\left(v,z^e_{\operatorname{clip}(t+\delta)}\right),
\label{eq:temporal_distance}
\end{equation}
where $\mathcal{V}$ is the speech-token vocabulary, $D_H$ denotes the coordinate-wise Hamming distance between the FSQ tuples associated with two tokens, and $w$ is the temporal
tolerance in token frames. The clipping operation restricts the evidence index to the valid sequence range. Using this distance, \csg{} selects the next token as
\begin{equation}
\vspace{-1mm}
\hat{z}_t
=
\arg\max_{v\in\mathcal{V}}
\left[
\log p_t(v)
-
\lambda D_{H,w}(v,z_t^e)
\right],
\label{eq:csg}
\vspace{-1.2mm}
\end{equation}
where $p_t(v)$ is the next-token probability predicted by Qwen-TSE conditioned on the speaker embedding, speaker-conditioned WavLM features, and previously selected tokens $\hat{z}_{<t}$. The parameter $\lambda$ balances model probability and agreement with the repaired evidence. Each selected token $\hat{z}_t$ is appended to the autoregressive history before the next decoding step. \textbf{Fixed CSG} uses $\lambda=1$ and same index evidence matching ($w=0$).

We additionally consider \textbf{adaptive \csg{}}, which sets the grounding
strength for each utterance according to its estimated difficulty.
Given the mixture $y$ and selected evidence $c^*$, we project $y$
onto $c^*$ and compute a projection-to-residual ratio:
\begin{equation}
\alpha^*
=
\left[
\frac{\langle y,c^*\rangle}
{\lVert c^*\rVert_2^2+\epsilon}
\right]_+,
\quad
\xi(y,c^*)
=
10\log_{10}
\frac{\lVert\alpha^*c^*\rVert_2^2+\epsilon}
{\lVert y-\alpha^*c^*\rVert_2^2+\epsilon}.
\label{eq:adaptive_difficulty}
\end{equation}
Here, $[x]_+=\max(0,x)$ and $\epsilon>0$ is a numerical stabilizer. The score $\xi$ compares the energy of the evidence-aligned component $\alpha^*c^*$ with that of the residual $y-\alpha^*c^*$. A fixed decreasing mapping $\lambda=g(\xi)$ applies stronger grounding at low
$\xi$ and weaker grounding at high $\xi$. Adaptive \csg{} uses $w=1$ (matching $t-1$, $t$, or $t+1$) and computes one $\lambda$ per utterance, which remains fixed during decoding. It uses only $y$ and $c^*$, without oracle information or model retraining. The generated speech-token sequence is then converted into the output waveform using the CosyVoice3 decoder~\cite{du2025cosyvoice3}.
\vspace{-6mm}
\section{Experimental Setup}\label{sec:exp}
\vspace{-2mm}
\subsection{Data}
\vspace{-1mm}
We use 16-kHz Libri2Mix-min mixtures~\cite{cosentino2020librimix}. Qwen-TSE model is post-trained on the training set with 13,900 mixtures evaluated in both target directions, yielding 27,800 trials. Clean development (DEV) and test set (TEST) each contain 3,000 mixtures and 6,000 trials. All enrollments are clean, same-speaker, different-utterance recordings. During training-data preparation, we use the clean target sources to identify and replace 29 invalid WeSep evidence waveforms. These source references are not used during DEV or TEST, where CDCS selects candidates
using only the target enrollment.

The noisy DEV and TEST sets use the official Libri2Mix/WHAM! mix-both condition \cite{wichern2019wham}, each containing 3,000 mixtures and 6,000 trials. We also construct 2,400 controlled trials per split from 240 mixtures at $-5$, $0$, $5$, $10$, and $15$ dB. The main results use natural-noise TEST together with controlled-SNR results. The main noisy evaluation pools 6,000 natural-noise and 2,400 controlled-SNR TEST trials, yielding 8,400 trials in total.
\vspace{-2mm}
\subsection{Model Configurations}
\vspace{-1mm}
CDCS and CSG are defined in Secs.~\ref{sec:cdcs} and~\ref{sec:grounding}. We use the released non-causal \texttt{spk\_emb\_100} and \texttt{tfmap\_context\_100} WeSep checkpoints~\cite{yu2023bsrnn,wang2024wesep}. Each uses a six-block, 128-dimensional TSE-BSRNN with a 512/128-sample STFT window/hop at 16~kHz. The global path uses a 192-dimensional utterance-level ECAPA embedding, whereas the TF-context path combines an enrollment TF map with 512-dimensional frame-level ECAPA features. All TSE and ECAPA parameters remain frozen. For the local enrollment views defined in Sec.~\ref{sec:cdcs}, we set $\tau=2$~s and use a 10-ms window-search hop. Each view therefore has duration
$\min(2~\mathrm{s},T_r/3)$, when $T_r<6$~s, the entire corresponding third is used without padding.

Qwen-TSE is initialized from Qwen2.5-0.5B-Instruct~\cite{qwen2024} with all parameters fully fine-tuned. We condition the WavLM encoder~\cite{chen2022wavlm} on the target-speaker embedding during mixture encoding. Layer-normalized linear projectors map the resulting 768-dimensional speaker-conditioned features and the 192-dimensional speaker embedding to Qwen's 896-dimensional hidden space. We train Qwen and the projectors for five epochs using teacher-forced cross-entropy, with the AdamW optimizer, an effective batch size of 8, and learning rates of $2\times10^{-5}$ and $10^{-4}$, respectively. We select the epoch-4 checkpoint, which achieves the lowest teacher-forced token cross-entropy on clean DEV.

At inference, all Qwen-based systems use greedy decoding for $\lceil25T_{\mathrm{mix}}\rceil$ steps, where $T_{\mathrm{mix}}$ is the duration of mixture $y$ in seconds. Fixed \csg{} uses $(\lambda,w)=(1,0)$, whereas adaptive \csg{} uses $w=1$. The mapping $g$ is selected on DEV and kept fixed for TEST.
\vspace{-2mm}
\subsection{Evaluation Metrics}
\vspace{-1mm}
All systems are evaluated using fixed metric models. A frozen Whisper-large-v3 recognizer~\cite{radford2023whisper} transcribes each output and the corresponding clean target and interferer sources. After text normalization, we report macro WER by averaging the
utterance-level WER of the output hypothesis against the target-source hypothesis. This measures ASR consistency rather than error against human transcripts. 

For a more reliable evaluation, we additionally report the \textbf{content-switch rate}. A content switch is counted when the output hypothesis has a lower WER against the interferer hypothesis than against the target hypothesis. Cases with equal WERs are not counted as content switches.

We also evaluate speaker-identity consistency using a frozen CosyVoice CAMPPlus
encoder $G$~\cite{wang2023camplusplus}, independent of the ECAPA encoder used by \cdcs{}. For an output $x$, we define the \textbf{speaker margin} $m_{\mathrm{spk}}(x)$ as
\begin{equation}
m_{\mathrm{spk}}(x)
=
\sims\!\left(G(x),G(s_{\mathrm{tar}})\right)
-
\sims\!\left(G(x),G(s_{\mathrm{int}})\right).
\vspace{-1.5mm}
\end{equation}
An \textbf{acoustic switch} is also counted when $m_{\mathrm{spk}}(x)<0$. The reported content- and acoustic-switch rates are the percentages of evaluation trials satisfying their respective criteria.

For perceptual speech-quality evaluation, we use DNSMOS P.808~\cite{reddy2021dnsmos} for both clean and noisy conditions, including the controlled-SNR experiments. We also report UTMOS~\cite{saeki2022utmos} as a complementary quality estimate. These quality scores are interpreted alongside separate measures of speaker identity and content consistency \cite{pirklbauer2023metrics}.

For the \cdcsFive{} ablation on deterministic TSE candidates, we
define the clean-reference SI-SDR margin~\cite{leroux2019sisdr} as
\begin{equation}
\Delta_{\mathrm{SI}}(x)
=
\mathrm{SI\mbox{-}SDR}(x,s_{\mathrm{tar}})
-
\mathrm{SI\mbox{-}SDR}(x,s_{\mathrm{int}}).
\vspace{-1.5mm}
\label{eq:sisdr_margin}
\end{equation}
A primary-extractor output is a high-confidence swap when $\Delta_{\mathrm{SI}}<-5$~dB and its interferer SI-SDR is positive. It is a primary-correct control when $\Delta_{\mathrm{SI}}\geq5$~dB and its target SI-SDR is positive. All other trials are excluded. A swap is recovered when $\Delta_{\mathrm{SI}}(c^*)>0$. For analysis only, an oracle uses clean references to select the candidate with the largest margin. These references are never used by \cdcs{}. Statistical analysis uses 10,000 paired bootstrap resamples and exact McNemar tests. We apply SI-SDR only to deterministic \cdcsFive{} candidates because generative outputs need not be sample-aligned with the reference.
\begin{table}[t]
\centering
\caption{Ablation study of CDCS. (a) Wrong-speaker error rates (\%, $\downarrow$), defined by $\Delta_{\mathrm{SI}} < 0$. Full/TF use the complete enrolment with global/TF-context conditioning, while E/M/L use local enrolment views with maximum duration $\tau$. (b) Recovery of high-confidence primary-confusion cases and new errors
on primary-correct controls. Oracle availability uses clean references
for analysis only.}
\label{tab:cdcs_mechanism}
\vspace{1mm}
{\fontsize{8pt}{8.4pt}\selectfont
\setlength{\tabcolsep}{1.5pt}

\renewcommand{\arraystretch}{0.94}
\vspace{1mm}
\textbf{(a) Wrong-speaker error rates}
\par\vspace{1mm}

\begin{tabular*}{\columnwidth}{
@{\extracolsep{\fill}}lcccc@{}
}
\toprule
& \multicolumn{2}{c}{\textbf{Clean} ($N{=}6{,}000$)}
& \multicolumn{2}{c}{\textbf{Noisy} ($N{=}8{,}400$)} \\
\cmidrule(lr){2-3}
\cmidrule(lr){4-5}

\textbf{Configuration}
& \textbf{DEV}
& \textbf{TEST}
& \textbf{DEV}
& \textbf{TEST} \\
\midrule

Full (no CDCS)
& 6.90 & 6.87
& 19.85 & 18.95 \\

Early (E)
& 8.52 & 8.57
& 20.61 & 19.96 \\

Middle (M)
& 8.27 & 8.90
& 20.92 & 20.73 \\

Late (L)
& 8.52 & 8.50
& 20.71 & 20.18 \\

TF-context (TF)
& 2.35 & 2.53
& 14.32 & 14.37 \\

\addlinespace[0.5mm]

Full + E/M/L
& 3.70 & 3.60
& 15.96 & 15.00 \\

Full + TF
& 1.17 & 1.18
& 12.00 & 11.30 \\

\textbf{CDCS-5}
& \textbf{0.88}
& \textbf{0.75}
& \textbf{11.02}
& \textbf{10.46} \\
\bottomrule
\end{tabular*}

\par\vspace{1mm}

\renewcommand{\arraystretch}{1.05}
\vspace{1mm}
\textbf{(b) Candidate recovery and selection safety}
\par\vspace{1mm}

\begin{tabular*}{\columnwidth}{
@{\extracolsep{\fill}}lccc@{}
}
\toprule
\textbf{Split}
& \textbf{CDCS-5 recovery}
& \textbf{Oracle availability}
& \textbf{New errors} \\
\midrule

Clean TEST
& 355/398 (89.20\%)
& 360/398 (90.45\%)
& 2/5,588 (0.04\%) \\

Noisy TEST
& 374/646 (57.89\%)
& 431/646 (66.72\%)
& 8/4,709 (0.17\%) \\
\bottomrule
\end{tabular*}
}

\vspace{-5mm}
\end{table}
\vspace{-4mm}
\section{Results and Discussion}
\label{sec:results}
\vspace{-2mm}
\subsection{System-Level Comparison}
\vspace{-1mm}
Table~\ref{tab:overall} compares five configurations. BSRNN-TSE is the deterministic full-enrollment baseline, while BSRNN-TSE-CDCS-5 directly returns the selected waveform $c^*$. Qwen-TSE-UD is the ungrounded generative baseline, and the fixed and adaptive CSG systems ground the same Qwen-TSE checkpoint using the CDCS-selected evidence.

\cdcsFive{} substantially improves the reliability of deterministic
TSE. On clean TEST, it reduces content and acoustic switching from
7.00\% and 6.98\% to 0.88\% and 0.78\%, respectively. Under noise, it
reduces WER from 39.85\% to 30.67\% and acoustic switching from 18.23\%
to 10.39\%. These results show that conditioning diversity can repair
speaker-confused estimates, although the direct output incurs a small
predicted-quality trade-off under severe noise.

Grounding provides complementary improvements for generative TSE.
On noisy TEST, fixed CSG reduces WER from 46.70\% to 42.11\%, content
switching from 12.79\% to 9.10\%, and acoustic switching from 1.68\%
to 0.51\%. Adaptive CSG achieves the strongest speaker margin and
predicted quality, whereas fixed CSG gives the lowest error rates.
This demonstrates that grounding suppresses speaker and content drift, with the two variants offering different reliability--quality
operating points.

Qwen-TSE has higher Whisper-based WER than direct \cdcsFive{} despite
its substantially better DNSMOS, UTMOS, and speaker consistency.
Case-level analysis suggests that some large WER degradations arise
from Whisper transcription errors, including hallucinated insertions,
rather than clear perceptual content failures. Such mismatches can be more pronounced under noise \cite{kim2024automatic} and for synthesized speech \cite{lau2023synthesizing}. We therefore treat Whisper-based WER as informative but not conclusive, and interpret it jointly with content switching, speaker consistency, and perceptual quality.
\vspace{-2mm}
\subsection{Ablation Study of \cdcsFive{}}
\vspace{-1mm}
Table~\ref{tab:cdcs_mechanism}(a) shows that TF-context provides the
strongest individual complement to the primary extractor, with
Full+TF clearly outperforming Full+E/M/L. Nevertheless, using all five candidates achieves the lowest wrong-speaker rates of 0.75\% on clean TEST and 10.46\% on noisy TEST. This indicates that the early, middle, and late views collectively recover failures missed by the two full-enrollment paths. Different temporal regions and context lengths may contain different amounts of speaker-discriminative information, making these views complementary despite their weaker individual performance.

Table~\ref{tab:cdcs_mechanism}(b) further shows that these improvements rarely corrupt originally correct outputs. \cdcsFive{} recovers 89.2\% of high-confidence confusions on clean TEST and 57.9\% under noise, while the new-error rate remains below 0.2\% in both conditions. Its performance is close to the oracle on clean speech, whereas the larger gap under noise suggests that severe corruption makes both candidate generation and selection more difficult. Table~\ref{tab:cdcs_mechanism} uses an SI-SDR-based error definition, so its values are not directly comparable with the encoder-based acoustic-switch rates in Table~\ref{tab:overall}.
\begin{figure}[t]
  \centering
  \includegraphics[width=0.96\columnwidth]{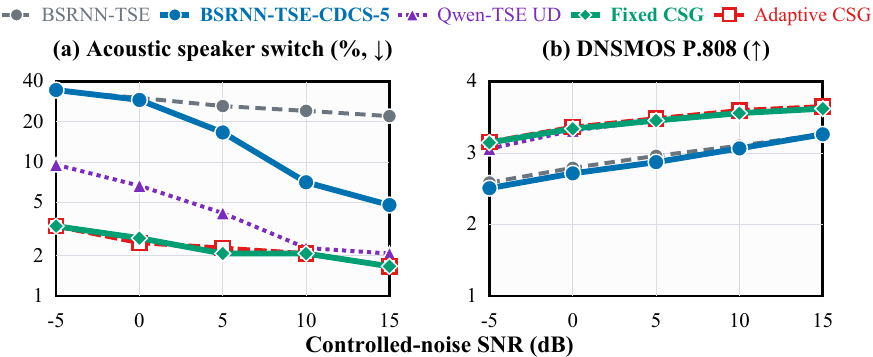}
  \vspace{-3.5mm}
  \caption{Controlled-noise TEST results across SNRs
  (480 target-conditioned trials per SNR): (a) acoustic-switch
  rate on a log scale and (b) DNSMOS P.808}
  \label{fig:snr_acoustic_dnsmos}
  \vspace{-5mm}
\end{figure}
\vspace{-2mm}
\subsection{Robustness across SNRs}
\vspace{-1mm}
Figure~\ref{fig:snr_acoustic_dnsmos} shows that all systems become
more reliable as SNR increases. Direct \cdcsFive{} increasingly
outperforms the primary extractor at moderate and high SNRs, whereas
both deterministic systems remain vulnerable at $-5$~dB. This trend
is consistent with the lower noisy-speech recovery in
Table~\ref{tab:cdcs_mechanism}. Across all SNRs, Qwen-TSE achieves substantially lower acoustic-switch rates and higher DNSMOS than the deterministic outputs. CSG provides its clearest gains over UD at low SNRs, showing that grounding helps suppress speaker drift under severe corruption. These results confirm the complementary roles of the two stages: \cdcsFive{} repairs the speaker identity of the evidence, while CSG preserves it during autoregressive generation.
\vspace{-2mm}
\section{Conclusion}
\vspace{-2mm}
We presented a repair-before-grounding framework that addresses two
distinct failure modes in LLM-based TSE: incorrect speaker evidence
and generative drift. Its key contribution, \cdcsFive{}, is a
training-free mechanism that constructs five complementary candidates
from frozen TSE paths and selects target-consistent evidence using only the mixture and target enrollment. The repaired waveform can be returned directly or used by CSG to constrain Qwen-TSE token generation. Across clean, noisy, and controlled-SNR conditions, \cdcsFive{} reduces speaker and content confusion with few regressions, while CSG further limits decoding drift and retains the quality advantage of generative resynthesis. Overall, evidence repair and decoding control provide complementary routes to more reliable generative TSE.

\vfill\pagebreak
\bibliographystyle{IEEEbib}
\bibliography{refs}

@article{zmolikova2023overview,
  author  = {K. Zmolikova and M. Delcroix and T. Ochiai and K. Kinoshita and J. {\v{C}}ernock{\'y} and D. Yu},
  title   = {Neural Target Speech Extraction: An Overview},
  journal = {IEEE Signal Process. Mag.},
  volume  = {40},
  number  = {3},
  pages   = {8--29},
  year    = {2023},
  doi     = {10.1109/MSP.2023.3240008}
}

@article{zmolikova2019speakerbeam,
  author  = {K. {\v{Z}}mol{\'i}kov{\'a} and others},
  title   = {{SpeakerBeam}: Speaker-Aware Neural Network for Target Speaker Extraction in Speech Mixtures},
  journal = {IEEE J. Sel. Topics Signal Process.},
  volume  = {13},
  number  = {4},
  pages   = {800--814},
  year    = {2019},
  doi     = {10.1109/JSTSP.2019.2922820}
}

@inproceedings{ge2020spexplus,
  author    = {M. Ge and C. Xu and L. Wang and E. S. Chng and J. Dang and H. Li},
  title     = {{SpEx+}: A Complete Time-Domain Speaker Extraction Network},
  booktitle = {Proc. Interspeech},
  pages     = {1406--1410},
  year      = {2020},
  doi       = {10.21437/Interspeech.2020-1397}
}

@inproceedings{sato2022enrolment,
  author    = {H. Sato and others},
  title     = {Strategies to Improve Robustness of Target Speech Extraction to Enrollment Variations},
  booktitle = {Proc. Interspeech},
  pages     = {996--1000},
  year      = {2022},
  doi       = {10.21437/Interspeech.2022-10767}
}

@inproceedings{zhao2022confusion,
  author    = {Z. Zhao and D. Yang and R. Gu and H. Zhang and Y. Zou},
  title     = {Target Confusion in End-to-End Speaker Extraction: Analysis and Approaches},
  booktitle = {Proc. Interspeech},
  pages     = {5333--5337},
  year      = {2022},
  doi       = {10.21437/Interspeech.2022-176}
}

@inproceedings{liu2023xsepformer,
  author    = {K. Liu and Z. Du and X. Wan and H. Zhou},
  title     = {{X-SEPFORMER}: End-to-End Speaker Extraction Network with Explicit Optimization on Speaker Confusion},
  booktitle = {Proc. ICASSP},
  pages     = {1--5},
  year      = {2023},
  doi       = {10.1109/ICASSP49357.2023.10095609}
}

@inproceedings{wang2024tcr,
  author    = {J. Wang and S. Wang and J. Li and K. Zhang and Y. Qian and H. Li},
  title     = {Enhancing Speaker Extraction Through Rectifying Target Confusion},
  booktitle = {Proc. IEEE Spoken Lang. Technol. Workshop (SLT)},
  pages     = {349--356},
  year      = {2024},
  doi       = {10.1109/SLT61566.2024.10832179}
}

@inproceedings{wang2024wesep,
  author    = {S. Wang and others},
  title     = {{WeSep}: A Scalable and Flexible Toolkit Towards Generalizable Target Speaker Extraction},
  booktitle = {Proc. Interspeech},
  pages     = {4273--4277},
  year      = {2024},
  doi       = {10.21437/Interspeech.2024-1840}
}

@inproceedings{yu2023bsrnn,
  author    = {J. Yu and H. Chen and Y. Luo and R. Gu and C. Weng},
  title     = {High-Fidelity Speech Enhancement with Band-Split {RNN}},
  booktitle = {Proc. Interspeech},
  pages     = {2483--2487},
  year      = {2023},
  doi       = {10.21437/Interspeech.2023-1433}
}

@inproceedings{yang2024context,
  author    = {X. Yang and C. Bao and J. Zhou and X. Chen},
  title     = {Target Speaker Extraction by Directly Exploiting Contextual Information in the Time-Frequency Domain},
  booktitle = {Proc. ICASSP},
  pages     = {10476--10480},
  year      = {2024},
  doi       = {10.1109/ICASSP48485.2024.10447529}
}

@inproceedings{desplanques2020ecapa,
  author    = {B. Desplanques and J. Thienpondt and K. Demuynck},
  title     = {{ECAPA-TDNN}: Emphasized Channel Attention, Propagation and Aggregation in {TDNN}-Based Speaker Verification},
  booktitle = {Proc. Interspeech},
  pages     = {3830--3834},
  year      = {2020},
  doi       = {10.21437/Interspeech.2020-2650}
}

@article{chen2022wavlm,
  author  = {S. Chen and others},
  title   = {{WavLM}: Large-Scale Self-Supervised Pre-Training for Full-Stack Speech Processing},
  journal = {IEEE J. Sel. Topics Signal Process.},
  volume  = {16},
  number  = {6},
  pages   = {1505--1518},
  year    = {2022},
  doi     = {10.1109/JSTSP.2022.3188113}
}

@incollection{tang2024tselm,
  author    = {B. Tang and B. Zeng and M. Li},
  title     = {{TSELM}: Target Speaker Extraction Using Discrete Tokens and Language Models},
  booktitle = {Man-Machine Speech Communication},
  pages     = {459--469},
  publisher = {Springer},
  address   = {Singapore},
  year      = {2026},
  doi       = {10.1007/978-981-95-5382-2_35}
}

@inproceedings{tang2025lauratse,
  author    = {B. Tang and B. Zeng and M. Li},
  title     = {{LauraTSE}: Target Speaker Extraction Using Auto-Regressive Decoder-Only Language Models},
  booktitle = {Proc. IEEE Autom. Speech Recognit. Understanding Workshop (ASRU)},
  pages     = {1--8},
  year      = {2025},
  doi       = {10.1109/ASRU65441.2025.11434620}
}

@inproceedings{kamo2023diffusion,
  author    = {N. Kamo and M. Delcroix and T. Nakatani},
  title     = {Target Speech Extraction with Conditional Diffusion Model},
  booktitle = {Proc. Interspeech},
  pages     = {176--180},
  year      = {2023},
  doi       = {10.21437/Interspeech.2023-1234}
}

@inproceedings{yu2024generation,
  author    = {L. Yu and W. Zhang and C. Du and L. Zhang and Z. Liang and Y. Qian},
  title     = {Generation-Based Target Speech Extraction with Speech Discretization and Vocoder},
  booktitle = {Proc. ICASSP},
  pages     = {12612--12616},
  year      = {2024},
  doi       = {10.1109/ICASSP48485.2024.10446418}
}

@article{zeng2026discgen,
  author        = {B. Zeng and B. Tang and W. Xiang and M. Li},
  title         = {Discriminative--Generative Target Speaker Extraction with Decoder-Only Language Models},
  journal       = {arXiv preprint arXiv:2601.06006},
  year          = {2026},
  eprint        = {2601.06006},
  archivePrefix = {arXiv},
  primaryClass  = {eess.AS},
  url           = {https://arxiv.org/abs/2601.06006}
}

@inproceedings{mentzer2024fsq,
  author    = {F. Mentzer and D. Minnen and E. Agustsson and M. Tschannen},
  title     = {Finite Scalar Quantization: {VQ-VAE} Made Simple},
  booktitle = {Proc. Int. Conf. Learn. Represent. (ICLR)},
  year      = {2024},
  url       = {https://proceedings.iclr.cc/paper_files/paper/2024/hash/e2dd53601de57c773343a7cdf09fae1c-Abstract-Conference.html}
}

@article{du2025cosyvoice3,
  author        = {Z. Du and others},
  title         = {{CosyVoice 3}: Towards In-the-Wild Speech Generation via Scaling-Up and Post-Training},
  journal       = {arXiv preprint arXiv:2505.17589},
  year          = {2025},
  eprint        = {2505.17589},
  archivePrefix = {arXiv},
  primaryClass  = {cs.SD},
  url           = {https://arxiv.org/abs/2505.17589}
}

@article{qwen2024,
  author        = {{Qwen Team}},
  title         = {{Qwen2.5} Technical Report},
  journal       = {arXiv preprint arXiv:2412.15115},
  year          = {2024},
  eprint        = {2412.15115},
  archivePrefix = {arXiv},
  primaryClass  = {cs.CL},
  url           = {https://arxiv.org/abs/2412.15115}
}

@article{cosentino2020librimix,
  author        = {J. Cosentino and M. Pariente and S. Cornell and A. Deleforge and E. Vincent},
  title         = {{LibriMix}: An Open-Source Dataset for Generalizable Speech Separation},
  journal       = {arXiv preprint arXiv:2005.11262},
  year          = {2020},
  eprint        = {2005.11262},
  archivePrefix = {arXiv},
  primaryClass  = {eess.AS},
  url           = {https://arxiv.org/abs/2005.11262}
}

@inproceedings{wichern2019wham,
  author    = {G. Wichern and others},
  title     = {{WHAM}!: Extending Speech Separation to Noisy Environments},
  booktitle = {Proc. Interspeech},
  pages     = {1368--1372},
  year      = {2019},
  doi       = {10.21437/Interspeech.2019-2821}
}

@inproceedings{radford2023whisper,
  author    = {A. Radford and J. W. Kim and T. Xu and G. Brockman and C. McLeavey and I. Sutskever},
  title     = {Robust Speech Recognition via Large-Scale Weak Supervision},
  booktitle = {Proc. Int. Conf. Mach. Learn. (ICML)},
  pages     = {28492--28518},
  year      = {2023},
  url       = {https://proceedings.mlr.press/v202/radford23a.html}
}

@inproceedings{leroux2019sisdr,
  author    = {J. Le Roux and S. Wisdom and H. Erdogan and J. R. Hershey},
  title     = {{SDR}---Half-Baked or Well Done?},
  booktitle = {Proc. ICASSP},
  pages     = {626--630},
  year      = {2019},
  doi       = {10.1109/ICASSP.2019.8683855}
}

@inproceedings{reddy2021dnsmos,
  author    = {C. K. A. Reddy and V. Gopal and R. Cutler},
  title     = {{DNSMOS}: A Non-Intrusive Perceptual Objective Speech Quality Metric to Evaluate Noise Suppressors},
  booktitle = {Proc. ICASSP},
  pages     = {6493--6497},
  year      = {2021},
  doi       = {10.1109/ICASSP39728.2021.9414878}
}

@inproceedings{saeki2022utmos,
  author    = {T. Saeki and D. Xin and W. Nakata and T. Koriyama and S. Takamichi and H. Saruwatari},
  title     = {{UTMOS}: {UTokyo-SaruLab} System for {VoiceMOS} Challenge 2022},
  booktitle = {Proc. Interspeech},
  pages     = {4521--4525},
  year      = {2022},
  doi       = {10.21437/Interspeech.2022-439}
}

@inproceedings{pirklbauer2023metrics,
  author    = {J. Pirklbauer and others},
  title     = {Evaluation Metrics for Generative Speech Enhancement Methods: Issues and Perspectives},
  booktitle = {Proc. 15th ITG Conf. Speech Commun.},
  pages     = {265--269},
  year      = {2023},
  publisher = {VDE},
  doi       = {10.30420/456164052}
}

@article{shi2026grounded,
  author        = {H. Shi and others},
  title         = {Grounded Decoding for Autoregressive Speech Enhancement via Adaptive Code-Space Grounding and Local {LLM} Refinement},
  journal       = {arXiv preprint arXiv:2609.04245},
  year          = {2026},
  eprint        = {2609.04245},
  archivePrefix = {arXiv},
  primaryClass  = {cs.SD},
  url           = {https://arxiv.org/abs/2609.04245}
}

@article{you2026multistep,
  author        = {Z. You and Y. Shi and L. Li and D. Wang},
  title         = {Training-Free Multi-Step Inference for Target Speaker Extraction},
  journal       = {arXiv preprint arXiv:2603.10921},
  year          = {2026},
  eprint        = {2603.10921},
  archivePrefix = {arXiv},
  primaryClass  = {eess.AS},
  url           = {https://arxiv.org/abs/2603.10921}
}

@article{liu2026evotse,
  author        = {Z. Liu and others},
  title         = {{EvoTSE}: Evolving Enrollment for Target Speaker Extraction},
  journal       = {arXiv preprint arXiv:2604.06810},
  year          = {2026},
  eprint        = {2604.06810},
  archivePrefix = {arXiv},
  primaryClass  = {eess.AS},
  url           = {https://arxiv.org/abs/2604.06810}
}

@inproceedings{wang2023camplusplus,
  author    = {H. Wang and S. Zheng and Y. Chen and L. Cheng and Q. Chen},
  title     = {{CAM++}: A Fast and Efficient Network for Speaker Verification Using Context-Aware Masking},
  booktitle = {Proc. Interspeech},
  pages     = {5301--5305},
  year      = {2023},
  doi       = {10.21437/Interspeech.2023-1513}
}

@article{kim2024automatic,
  author  = {S.-E. Kim and B. R. Chernyak and O. Seleznova and J. Keshet and M. Goldrick and A. R. Bradlow},
  title   = {Automatic Recognition of Second Language Speech-in-Noise},
  journal = {JASA Express Lett.},
  volume  = {4},
  number  = {2},
  year    = {2024},
  note    = {Art. no. 025204},
  doi     = {10.1121/10.0024877}
}

@inproceedings{lau2023synthesizing,
  author    = {J. K. Lau and others},
  title     = {Synthesizing Speech Test Cases with Text-to-Speech? {An} Empirical Study on the False Alarms in Automated Speech Recognition Testing},
  booktitle = {Proc. 32nd ACM SIGSOFT Int. Symp. Softw. Test. Anal. (ISSTA)},
  pages     = {1169--1181},
  year      = {2023},
  doi       = {10.1145/3597926.3598126}
}

\end{document}